\documentclass[twocolumn]{autart_arxiv}    

\usepackage{graphicx}          
\usepackage{wrapfig}
\usepackage{amsmath} 
\usepackage{amssymb}  
\usepackage[dvipsnames]{xcolor}
\usepackage{url}
\usepackage{natbib}

\DeclareMathOperator{\diag}{diag}

\DeclareMathOperator{\rank}{rank}
\DeclareMathOperator{\Span}{span}
\DeclareMathOperator{\Null}{Null}
\DeclareMathOperator{\blkdiag}{blkdiag}
\DeclareMathOperator{\tr}{tr}
\DeclareMathOperator{\sgn}{sgn}

\newtheorem{lemma}{Lemma}
\newtheorem{assumption}{Assumption}
\newtheorem{remark}{Remark}
\newtheorem{definition}{Definition}

\newtheorem{proposition}{Proposition}

\begin{document}
	
	\begin{frontmatter}
		\vspace{-0.3cm}
		\title{Relaxed bearing rigidity and bearing formation control under persistence of excitation\thanksref{footnoteinfo}} 
		
		\thanks[footnoteinfo]{This paper was partially presented at the 59th IEEE Conference on Decision and Control (CDC), 2020. Corresponding author Zhiqi Tang. }
		\vspace{-0.6cm}
		\author[lisbon,Nice]{Zhiqi Tang}\ead{zhiqitang@tecnico.ulisboa.pt},    
		\author[lisbon]{Rita Cunha}\ead{rita@isr.tecnico.ulisboa.pt},               
		\author[Nice,IUF]{Tarek Hamel}\ead{ thamel@i3s.unice.fr},
		\author[macau,lisbon]{Carlos Silvestre}\ead{ csilvestre@umac.mo}
		\address[lisbon]{ISR, IST, Universidade de Lisboa, Portugal. }  
		\address[Nice]{I3S-CNRS, Universit\'{e} C\^{o}te d'Azur, Nice-Sophia Antipolis, France.}             
		\address[IUF]{IUF, Institut Universitaire de France.}        
		\address[macau]{Faculty of Science and Technology of the University of Macau, Macao, China.}
		\vspace{-0.3cm}
		\vspace{-0.6cm}
		\begin{abstract}                          
			This paper addresses the problem of time-varying bearing formation control in $d$ $(d\ge 2)$-dimensional Euclidean space by exploring \textit{Persistence of Excitation} (PE) of the desired bearing reference. A general concept of \textit{Bearing Persistently Exciting} (BPE) formation defined in $d$-dimensional space is here fully developed.
			By providing a desired formation that is BPE, distributed control laws for multi-agent systems under both single- and double-integrator dynamics are proposed using bearing measurements (along with velocity measurements when the agents are described by double-integrator dynamics), which guarantee uniform exponential stabilization of the desired formation in terms of shape and scale. A key contribution of this work is to show that the classical bearing rigidity condition on the graph topology, required for achieving the stabilization of a formation up to a scaling factor, is relaxed and extended in a natural manner by exploring PE conditions imposed either on a specific set of desired bearing vectors or on the whole desired formation. Simulation results are provided to illustrate the performance of the proposed control method.
		\end{abstract}
	\end{frontmatter}
	\section{INTRODUCTION}
		\vspace{-0.2cm}
	Bearing formation control has received growing attention in both the robotics and control communities due to its minimal requirements on the sensing ability of each agent. Early works on bearing-based formation control were mainly focused on controlling the subtended bearing angles that are measured in each agent's local coordinate frame  and were limited to planar formations only (\cite{basiri2010distributed,bishop2011very}). The main body of work however builds on the concept of bearing rigidity theory (also termed parallel rigidity) e.g. \cite{servatius1999constraining,eren2003sensor,zhao2016bearing}, which investigates the conditions for which a static formation is uniquely determined up to a translation and a scale factor given the corresponding constant bearing measurements. Under the assumption that the desired formation is Infinitesimally Bearing Rigid (IBR), the work \cite{zhao2016bearing} proposes a bearing-only controller that guarantees convergence to the target formation up to a scale factor and a translation. To remove the scale ambiguity, it is still necessary to have at least two leaders or one known distance between a pair of agents (e.g. \cite{zhao2019bearing}).
	In multi-agent systems, minimal communication among agents is always advantageous in terms of power consumption and important to determine tolerable connection losses. 
	Hence, minimal bearing rigidity, which determines whether or not the connections in a graph are minimal in the sense that removing any of these connections will result in loosing bearing rigidity, has been studied in \cite{eren2003sensor} and \cite{trinh2019minimal}.
	
	The concept of bearing rigidity explored in the literature is mainly focused on static bearing references. However, the natural behavior of multi-agent formations typically evolves in time and requires dynamic coordination among agents, such as in fish schooling or bird flocking. This draws our interests to time-varying bearing formations and to the well-known concept of PE,  which has been recently exploited only for relative position estimation in a bearing-based circumnavigation task in \cite{shao2018multi}. 
	Inspired by \cite{le2017observers,hamel2017position},  we introduced in \cite{tang2020leader,tang2021formation} the concept of BPE formation and \textit{Relaxed Bearing Rigid} (RBR) formation for a leader-follower structure, which loosens the constraints imposed on the graph topology required by the leader-first follower structure defined in \cite{trinh2019bearing}. Additionally, we proposed leader-follower bearing control laws that achieve exponential stabilization of the formation tracking error in terms of position and velocity, provided the desired formation is BPE.
	\begin{figure}[!t]
		\centering
		\includegraphics[width=2.6in]{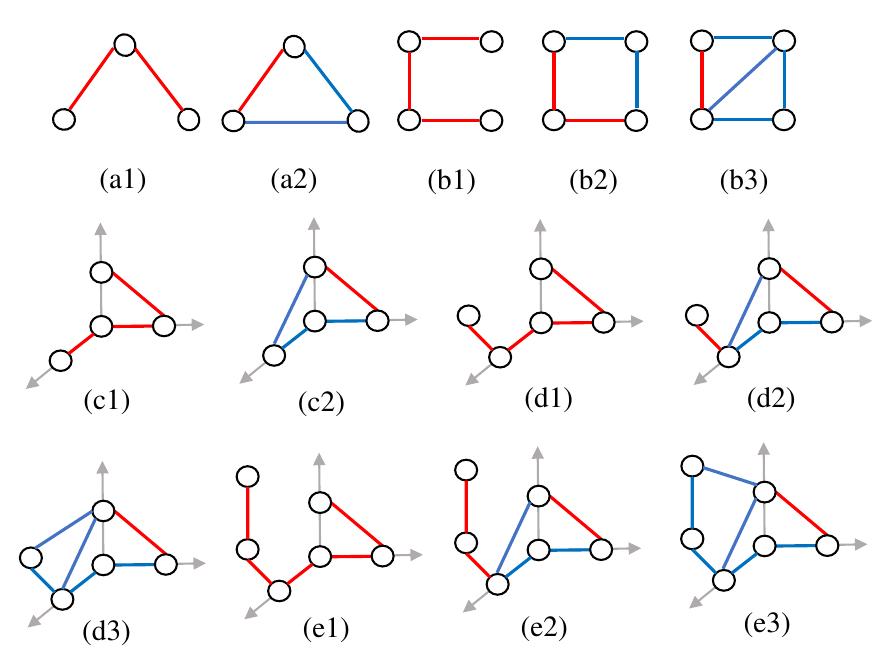}	
		\caption{Examples of BPE formations in two $(a1-b3)$ and three-dimensional space $(c1-e3)$. Red lines represent edges for which the corresponding bearing vector are PE and blue lines represent edges for which the corresponding bearing vectors are not necessarily PE. }
		\label{fig:pe_edges3D}
	\end{figure}
	
	This paper presents a coherent generalization of our previous solutions (\cite{tang2021formation,tang2020bearing}) to formations under general undirected graph topologies and fully develops the general concept of BPE formation defined in $d\ (d\ge 2)$-dimensional Euclidean space, whose configuration can be uniquely determined up to a translation using only bearing and velocity measurements.
	We provide necessary conditions for having a BPE formation when the topology lacks the connections required by IBR formation, 
	and necessary and sufficient conditions that ensure BPE for three particular cases, including the case of vertex addition. We also define a particular subclass of BPE formations called RBR formations, which guarantee uniqueness of a geometric shape and its scale through a continuous similarity transformation  
	involving a time-varying rotation 
	without imposing 
	the classical bearing rigid conditions on the graph topology. For example, the BPE formations shown in Fig.~\ref{fig:pe_edges3D}-\scalebox{0.8}{$(a1),(b1-2),(c1-2),(d1-2)$} and \scalebox{0.8}{$(e1-2)$} are not IBR but can be RBR.
	Based on these results, distributed control laws are proposed for a multi-agent system (with both single- or double-integrator dynamics) to track a BPE desired formation using only bearing measurements (along with velocity measurements for double-integrator dynamics), which achieve uniform exponential (UE) stabilization of the formation to the desired one up to a translation Euclidean vector, under any undirected graph that has a spanning tree.
	A safe set of initial conditions that guarantees collision avoidance during transient is also provided.
	
	The body of the paper is organized as follows. Section \ref{sec:graph} presents mathematical background on graph theory and formation control. Section \ref{sec:BPE formation} introduces the BPE theory and the definition of RBR formation. Section \ref{sec:control} and \ref{sec:control2} presents the proposed bearing formation control laws along with stability analysis for both single- and double-integrator dynamics, respectively. Section \ref{sec:sim} presents simulation results obtained with the proposed control strategy. The paper concludes with some final comments in Section \ref{sec:conc}.
	\section{Preliminaries}\label{sec:graph}
	Let $\mathbb{S}^{d-1}:=\{y\in\mathbb{R}^d:\|y\|=1\}$ denote the $(d-1)$-Sphere ($d \geq 2$) and $\|.\|$ the Euclidean norm. The null space, rank, trace and determinant of a matrix are denoted by $\Null(.)$, $\rank(.)$, $\tr(.)$ and $\det(.)$, respectively. For any positive symmetric matrix of dimension $n\times n$, $\lambda_{\max}(.)(\lambda_{\min}(.))$ represents the maximum (minimum) eigenvalue of its matrix argument. Let $\mathrm{mod}(a,b)$ denote the remainder of $a/b$, with $a\in \mathbb N$ and  $b\in \mathbb N^*$. The signum function is denoted by $\sgn(.)$. For any $r\in\mathbb R^{+}$, $\llcorner r \lrcorner $ denotes the integer part of $r$. The matrix $I_d$ represents the identity matrix of dimension $d\times d$. The matrix $0_d$ and $0_{d\times q}$ represents the zero matrix of dimension $d\times d$ and $d\times q$, 
	respectively. The operator $\otimes$ denotes the Kronecker product, $\boldsymbol 1_n = [1,\ldots,1]^T \in \mathbb R^{n}$ the column vector of ones, $\boldsymbol 0_n\in \mathbb R^{n}$ the column vector of zeros and $\diag(A_i) = \blkdiag \{A_1, \ldots, A_n\} \in \mathbb{R}^{nd\times nd}$ the block diagonal matrix with elements given by $A_i\in \mathbb{R}^{d\times d}$ for $i=1,\ldots,n$.
	For any $y\in \mathbb{S}^{d-1}$, we define:
	$\pi_y := I_d - y y^{\top} \geq 0$ as the orthogonal projection operator in $\mathbb{R}^d$ onto the $(d-1)$-dimensional vector subspace orthogonal to $y$.
	\subsection{Graph theory}\label{sec:LF}
	Consider a system of $n$ connected agents. The underlying interaction topology can be modeled as an undirected graph $\mathcal{G} := (\mathcal{V}, \mathcal{E})$, where $\mathcal{V}=\{1,\ldots,n\}$ ($n \geq 2$) is the set of vertices and $\mathcal{E} \subseteq \mathcal{V} \times \mathcal{V}$ is the set of undirected edges. Two vertices $i$ and $j$ are called adjacent (or neighbors)
	when $\{i,j\}\in \mathcal{E}$. The set of neighbors of agent $i$ is denoted by $\mathcal{N}_i:=\{j\in\mathcal{V}|\{i,j\}\in\mathcal{E}\}$. If $j\in\mathcal N_i$, it follows that $i\in \mathcal N_j$, since the edge set in an undirected graph consists of unordered vertex pairs. Let $m=|\mathcal E|$ be the cardinality of the set $\mathcal E$. A graph $\mathcal G$ is connected if there exists a path between every pair of vertices in $\mathcal G$ and in that case $m \geq n-1$. A graph $\mathcal G$ is said to be acyclic if it has no circuits. 
	 A spanning tree of a graph $\mathcal G$ is a connected acyclic subgraph of $\mathcal G$ involving all the vertices of $\mathcal G$. If the graph is acyclic and has a spanning tree, $m=n-1$.
	An oriented graph is an undirected graph together with an orientation that assigns a direction to each edge. The incidence matrix $H\in \mathbb{R}^{m\times n}$ of an oriented graph is the $\{0,\pm 1\}$-matrix with rows indexed by edges and columns by vertices: $[H]_{ki}=1$ if vertex $i$ is the head of the edge $k$, $[H]_{ki}=-1$ if it is the tail, and $[H]_{ki}=0$ otherwise, implying that $H\boldsymbol 1_n=0$. For a connected graph, or equivalently a graph having a spanning tree, one always has rank$(H)=n-1$.
	\subsection{Formation control} \label{subsection:formation control}
	Consider an undirected graph $\mathcal{G}=(\mathcal{V},\mathcal{E})$, let $p_i \in \mathbb{R}^d$  and $v_i \in \mathbb{R}^d$, $(d\ge 2)$ denote the position and velocity, respectively, of each agent $i\in\mathcal{V}$ both expressed in a common inertial frame. Then, the stacked vector $\boldsymbol{p}=[p_1^\top,...,p_n^\top]^\top\in \mathbb{R}^{dn}$ ($n\geq2$) is called a configuration of $\mathcal{G}$. The graph $\mathcal{G}$ and the configuration $\boldsymbol{p}$ together define a formation $\mathcal{G}(\boldsymbol p)$ in the $d$-dimensional Euclidean space. Let $\boldsymbol v:=\dot {\boldsymbol p}=[v_1^\top,\ldots,v_n^\top]^\top\in \mathbb R ^{dn}$. For any formation, we define the relative position
	\begin{equation}\label{eq:pij}
	p_{ij}:=p_j-p_i,\ \{i,j\}\in\mathcal E.
	\end{equation}
	As long as $\|p_{ij}\|\ne 0$, the bearing of agent $j$ relative to agent $i$ is given by the unit vector
	\begin{equation}
	\label{eq:gij}
	g_{ij} := p_{ij}/\|p_{ij}\| \in \mathbb{S}^{d-1}.
	\end{equation}
	Consider an arbitrary orientation of the graph and denote
	$$\bar p_k:= p_{ij},\ k\in \{1,\ldots,m\},$$
	as the edge vector with assigned direction such that $i$ and $j$ are, respectively, the initial and the
	terminal nodes of $\bar p_k$. Denote the corresponding bearing vector by
	$$\bar g_k:= \frac{\bar p_k}{\|\bar p_k\|}\in \mathbb{S}^{d-1},\ k\in \{1,\ldots,m\}.$$
 For any formation control problem involving relative position measurements, the graph Laplacian matrix
 \begin{equation} \label{eq:Laplacian}\scalebox{0.95}{$
 L:=\bar H^\top \bar H, \text{ with } \bar H=H \otimes I_d$}
 \end{equation}
  is adopted in any distributed control law design aiming to drive the configuration $\boldsymbol p$ to the desired one up to translation Euclidean vector (\cite{mesbahi2010graph,oh2015survey}).	
If the graph is connected, one has $\rank(L)=dn-d$, $\Null(L)=\Span\{U\}$, with $U=\boldsymbol 1_n\otimes I_d$ and hence, by adopting $\lambda_i$ as the $i$th eigenvalue of $L$ under a non-increasing order, one ensures that $\lambda_{dn-d}$ is the smallest positive eigenvalue of $L$. 

 For bearing-based formation control problems, the bearing Laplacian matrix is defined as
	\begin{equation}\label{eq:bearingL}\scalebox{0.95}{$
	L_B(\boldsymbol p(t)):=\bar H^\top \Pi\bar H \text{, with } \Pi=\diag(\pi_{\bar g_k}).$}
	\end{equation}
	Since $\Span\{U,\boldsymbol p\}\subset \Null(L_B)$ it follows that $\rank(L_B)\le dn-d-1$. According to \cite{zhao2016bearing} (in which only constant bearing are considered), if the formation is IBR then $\rank(L_B)= dn-d-1$,  $\Null(L_B)=\Span\{U,\boldsymbol p\}$ and $\bar m\le m\le n(n-1)/2$, where
	\begin{equation}\label{eq:f(n,d)}\scalebox{0.9}{$
	\bar m(n,d)=\left\{
	\begin{aligned}
	&n, \ n\le d+1\\
	&1+\llcorner \frac{n-2}{d-1}\lrcorner \times d+\mathrm{mod}(n-2,d-1)\\&+\sgn(\mathrm{mod}(n-2,d-1)), \ n\ge d+1,
	\end{aligned}\right.$}
	\end{equation}
	is the minimal number of edges that guarantees $\rank(L_B(\boldsymbol p(t)))=dn-d-1$, \cite{trinh2019minimal}.
\section {Bearing persistence of excitation in $\mathbb{R}^d$}\label{sec:BPE formation}
After a short presentation of the persistence of excitation and persistently exciting bearing (Subsection \ref{sec:PersistentofExcitation}),  we formalize in Subsection \ref{sec:BearingFomation} the concept of \textit{Bearing Persistently Exciting} (BPE) formation under which the formation's configuration can be uniquely determined up to a translational Euclidean vector using only bearing and velocity measurements. Then we provide some results on necessary conditions, and necessary $\&$  sufficient conditions to guarantee the BPE formation. Finally we introduce the concept of RBR formation as a particular situation of BPE formation.
\subsection{Persistence of excitation}\label{sec:PersistentofExcitation}
\begin{definition}
	A positive semi-definite matrix $\Sigma(t)\in \mathbb{R}^{n\times n}$, is called \textit{persistently exciting} (PE) if there exist $T>0$ and $\mu>0$ such that for all $t\geq0$
	\begin{equation}\scalebox{0.95}{$
	\frac 1 T \int_{t}^{t+T}\Sigma(\tau)d\tau\ge\mu I. \label{eq:pe}$}
	\end{equation}
	\label{def:pe of matrix}
\end{definition}
\begin{definition}\label{def:pe}
	A direction $y(t)\in \mathbb{S}^{d-1}$, is called PE if the matrix $\pi_{y(t)}$ satisfies the PE condition from Definition \ref{def:pe of matrix}.
\end{definition}
\begin{lemma}\label{lem:pe_norm}
	Assume that $y(t)\in \mathbb{S}^{d-1}$ and $\dot y(t)$ is uniformly continuous, then relation \eqref{eq:pe} with $\Sigma(\tau)=\pi_{y(\tau)}$ is equivalent to:
	$\forall t\geq0$, there exist $(T,\epsilon)>0$ and $\tau \in [t,\ t+T]$ 
	such that $\|\dot y(\tau)\|\ge \epsilon$ . 
\end{lemma}
\begin{lemma}\label{lem:Q_norm}
	Let $Q:=\sum\limits_{i=1}^{l}\pi_{y_i}$. The matrix $Q$ is PE, if one of the following conditions is satisfied:
	\begin{enumerate}
		\item there is at least one PE direction $y_i$,
		\item there are at least two uniformly non-collinear directions $y_{i}$ and $y_j$, $i,j\in\{1,...,l\},\ i\ne j$. That is:
		$\forall t\ge 0, \; \exists \epsilon_1>0$ such that $|y_i(t)^\top y_j(t)|\le 1-\epsilon_1$.
	\end{enumerate}
\end{lemma}The proofs of Lemmas \ref{lem:pe_norm} and \ref{lem:Q_norm} are given in \cite{le2017observers}.
\subsection{BPE formation and Relaxed Bearing Rigidity}\label{sec:BearingFomation}
For any formation $\mathcal G (\boldsymbol p(t))$ defined in $\mathbb R ^{d}$ with Laplacian $L$ and Bearing Laplacian $L_B$ given by \eqref{eq:Laplacian} and \eqref{eq:bearingL}, respectively, we define a BPE formation as follows.
\begin{definition}\label{def:pe of LB}
		A formation $\mathcal G(\boldsymbol p(t))$ is \textit{Bearing Persistently Exciting} (BPE) if $\mathcal G$ has a spanning tree and its bearing Laplacian matrix is PE:
		\begin{equation}
		\forall t\geq0, \exists T>0, \mu>0, \ \frac 1 T \int_{t}^{t+T}L_B(\boldsymbol p(\tau))d\tau\ge\mu L. \label{eq:pe_LB}
		\end{equation}
\end{definition}
Note that, the PE condition for the bearing Laplacian introduced in Definition \ref{def:pe of LB} is less restrictive than the PE condition on the bearing matrix $\Pi$ in \eqref{eq:bearingL} from Definition \ref{def:pe of matrix}. In particular, having a matrix $\Pi$ that is PE is sufficient but not necessary 
to ensure that $L_B = \bar{H}^\top\Pi\bar{H}$ is also PE in the sense of \eqref{eq:pe_LB}.
The following Theorem proposes an observer for the configuration of a formation (using only bearing and velocity measurements) that ensures global UE convergence of the observer error to a specific constant translation Euclidean vector in $\mathbb R^{dn}$, provided the formation is BPE.  

\begin{thm}\label{thm:shape}
	Consider a formation $\mathcal G (\boldsymbol p(t))$ defined in $\mathbb R ^{d}$. Assume that the bearing measurements $\{g_k\}_{ k\in\{1\ldots m\}}$ under an arbitrary orientation of the graph and the velocity measurements $\{v_i\}_{ i\in \{1 \ldots n\}}$ are well-defined, bounded and known. Let $\hat {\boldsymbol p} \in \mathbb{R}^{dn}$  denote the estimate of $\boldsymbol p$ with dynamics:
	\begin{equation} \label{eq:observer}
	\begin{aligned}
	\dot{\hat {\boldsymbol p}}=\boldsymbol v-kL_B(\boldsymbol p(t))\hat {\boldsymbol p},\;\; k>0,\; \hat {\boldsymbol p}(0)= \hat {\boldsymbol p}_0
	\end{aligned}
	\end{equation}
If $\mathcal G (p(t))$ is BPE, then for any initial condition the estimated configuration $\hat{\boldsymbol  p}(t)$ converges globally UE to $\boldsymbol p(t)+U\xi_0$, with $\xi_0$ a constant translational vector in $\mathbb R ^{d}$ defining the relative error between the estimated centroid and the actual one. 
\end{thm}
 	\begin{pf}
 	Define the relative error $ \xi_0(t):=\frac{1}{n}U^\top(\hat{\boldsymbol p}(t)-\boldsymbol p(t))\in \mathbb R^d$ and recall that $U=\boldsymbol 1_n\otimes I_d$, $U^\top L_B = 0$, and $L_B U = 0$. One can verify that  $\dot\xi_0\equiv 0$, $\xi_0$ is constant and hence $\xi_0(t)= \frac{1}{n}U^\top (\hat{\boldsymbol p}(0)-\boldsymbol p(0))$. Consider the error variable $\boldsymbol \zeta(t)$ defined such that $\hat{\boldsymbol p}(t)-\boldsymbol p(t) = \boldsymbol \zeta(t) + U \xi_0$ and $\boldsymbol \zeta(t)$ and $U\xi_0$ are orthogonal. Then, the corresponding dynamics can be obtained from \eqref{eq:observer}:
 	\begin{equation} \label{eq:observer_dynamics}
 	\dot{\boldsymbol \zeta}=-kL_B(\boldsymbol p(t)) \boldsymbol \zeta.
 	\end{equation}
 	Since the formation is BPE, $\forall x\in \mathbb{R}^{dn-d}$ satisfying $U^\top x\equiv0$, there exists a $T>0$ and $0<\mu <1$ such that, $\forall t$, $\frac 1 T x^\top \int_t^{t+T}L_B(\boldsymbol p(\tau))d \tau x\ge \mu x^\top \bar H^\top \bar Hx\ge \mu \lambda_{dn-d}\|x\|^2$, where $\lambda_{dn-d}>0$ is the smallest positive eigenvalue of $\bar H^\top \bar H$ (see Sect. \ref{subsection:formation control}). Using similar arguments as in the proof of \cite[Lemma 5]{loria2002uniform}, one can ensure that the equilibrium $\boldsymbol \zeta=0$ is uniformly globally exponentially (UGE) stable. Therefore, one concludes that $\hat{\boldsymbol p}$ converges UGE to the unique $\boldsymbol p$ up to a translational vector $U\xi_0$.
 	%
 \end{pf}
 Now, in order to explore the properties of BPE formations, the next Lemma extends Theorem 4.1 of \cite{trinh2019minimal} to provide a necessary condition on the number of PE bearings for having a BPE formation 
when $(n-1) \le m < \bar{m}$ (e.g. Fig. \ref{fig:pe_edges3D}-$(a1),(b1),(b2),(c1),(d1),(d2),(e1)$ and $(e2)$).
\begin{lemma}
	Consider a formation $\mathcal G (\boldsymbol p(t))$ defined in $\mathbb R^{d}$ involving $n$ agents and $m$ edges.
If i) the formation is BPE; and ii) 
$(n-1)\le m <\bar{m}$ (where $\bar{m}$ the minimal number of edges that guarantees $\rank(L_B(\boldsymbol p(t)))=dn-d-1$ defined in \eqref{eq:f(n,d)}), then $(d-1)(n-1)\le\rank(L_B(\boldsymbol p(t)))<dn-d-1, \ \forall t\ge 0$ and
the number of PE bearing vectors inside the formation, $m_{PE}$, satisfies  $ m_{PE} \geq d(n-1)-(d-1)m$.
\end{lemma}
 \begin{pf}
 		Since the formation is BPE, there is a spanning tree in $\mathcal G$ and inequality \eqref{eq:pe_LB} is satisfied. Due to the fact that $(n-1)\le m <\bar{m}$, it is obvious to conclude that $(d-1)(n-1)\le\rank(L_B(\boldsymbol p(t)))<dn-d-1, \ \forall t\ge 0$. Inequality \eqref{eq:pe_LB} implies that there exist $\mu>0$ and $ T>0$, $\forall t\ge0$ and $\forall \boldsymbol x\in R^{dn}$ such that $\bar{H}\boldsymbol x \neq 0$, we have $\frac 1 T \boldsymbol x^\top \int_t^{t+ T}L_B(\boldsymbol p(\tau))d\tau\boldsymbol  x\ge  \mu \boldsymbol x^\top L\boldsymbol x$ or equivalently $\frac 1 T \boldsymbol w^\top\int_t^{t+ T}\Pi(\tau)d\tau \boldsymbol w\ge \mu \|\boldsymbol w\|^2$, with $\boldsymbol w=\bar{H} \boldsymbol x\in \mathbb R^{dm}$. 
		
 		We proceed the remaining proof by contradiction. Assume that $ m_{PE}\le d(n-1)-(d-1)m-1$.
 		Since we have $m-m_{PE}$ non-PE bearings and for each non-PE bearing $\bar g_k$ there is a
 		$\lambda_{\min}(\frac 1 T\int_{t}^{t+T}\pi_{\bar g_k(\tau)}d\tau)<\mu$, it is straightforward to verify that $\lambda_{dn-d}(\frac 1 T\int_t^{t+T}\Pi(\tau)d\tau)\le \lambda_{dm-(m- m_{PE})+1}(\frac 1 T\int_t^{t+T}\Pi(\tau)d\tau)< \mu$ (where $\lambda_i(.)$ represents the $i$th eigenvalue of a symmetric matrix under a non-increasing order).
		
 		Now, using the fact that $\rank(\bar H)=dn-d$, we can ensure that if $\boldsymbol x=(x_1^\top, \ldots,x_n^\top)$ has $dn$ independent entries (each $x_i \in \mathbb{R}^{d}$), then there exists a $\boldsymbol w=\bar H \boldsymbol x$ with $dn-d$ independent entries such that $\frac 1 T \boldsymbol w^\top\int_t^{t+T}\Pi(\tau)d\tau \boldsymbol w <\mu \|\boldsymbol w\|^2$, which yields a contradiction.
 \end{pf}
To complement the above result, the following Theorem provides necessary and sufficient conditions on the PE bearings that ensure a BPE formation for three particular cases. 
\begin{thm}\label{thm:min_pe}
	Consider a formation $\mathcal G (\boldsymbol p(t))$ defined in $\mathbb R^{d}$ with vertex set $\mathcal V$ ($|\mathcal V|=n$) and edge set $\mathcal E$ ($|\mathcal E|=m$). The formation is BPE if and only if any of the following applies:
	\begin{enumerate}
		\item all bearings are PE, i.e. $\bar g_k(t)$ satisfies the PE condition for all $k\in\{1,\ldots,m\}$, when the graph $\mathcal G$ is acyclic and has a spanning tree ($m=n-1$ and $\rank(L_B(\boldsymbol p(t)))=(d-1)(n-1), \ \forall t\ge0$);
		\item at least one bearing $\bar g_k,\ k\in\{1,\ldots,m\}$ is PE, when $\mathcal G(\boldsymbol p(t))$ is IBR ($\bar{m}\leq m \leq n(n-1)/2$ and $\rank(L_B(\boldsymbol p(t)))=dn-d-1, \ \forall t\ge0$);
		\item $\sum_{k=m'+1} ^{m}\pi_{\bar{g}_k}$ is PE when $\mathcal G(\boldsymbol p(t)) \in \mathbb R^d$ is designed by adding a new agent to a BPE formation $\mathcal G'(\boldsymbol p'(t))$, 
		with vertex set $\mathcal V ^{'}$ and edge set $\mathcal E^{'}$, such that $\mathcal V=\mathcal V^{'}\cup \{l\}$, $\mathcal E^{'}\subset \mathcal E$, and $|\mathcal E^{'}|=m^{'}$.
	\end{enumerate}
\end{thm}
 \begin{pf}
 	Since in the three particular cases the graph $\mathcal G$ is connected (has a spanning tree), proof of BPE formation is equivalent to show that the bearing Laplacian $\mathcal L_B(\boldsymbol p(t))=\bar H^\top \Pi(\boldsymbol p(t))\bar H$ is PE.
	
 	Proof of Item (1):
 	If $\bar g_k(t)$ satisfies the PE condition  $\forall k=\{1,\ldots,m\}$, this implies that the matrix $\Pi(t)$ is PE and hence it is obvious to conclude that $L_B(\boldsymbol p(t))$ is PE.
 	Conversely, if $L_B(\boldsymbol p(t))$ is PE then there exist $T>0$ and $\mu >0$ such that, $\forall t\ge0$,  $\frac 1 T\int_{t}^{t+T}L_B(\boldsymbol p(\tau))d\tau\ge\mu L$. Now, since the $\bar{H}$ is a constant matrix with $\rank(\bar{H})=d(n-1)$ and $\Pi(t) \in \mathbb{R}^{d(n-1)\times d(n-1)}$ it follows that $\Pi(t) \in \mathbb{R}^{d(n-1)\times d(n-1)}$ should satisfy the PE condition in equation \eqref{eq:pe}.  This in turn implies that each $\bar g_k(t)$ satisfies the PE condition in Definition \ref{def:pe},  $\forall k\in\{1,\ldots,n-1\}$.
	
 	Proof of Item (2):
 Let $S=\{\mathring {\boldsymbol p} \in S| \mathring {\boldsymbol p}=[\mathring p_1^\top,\ldots,\mathring p_n^\top]\in\mathbb R^{dn}\}$ be the set of all possible fixed configurations under the formation $\mathcal G(\mathring {\boldsymbol p})$ leading to $\rank(L_B(\mathring {\boldsymbol p}))=dn-d-1$. 
 	This in turn implies that for any $\boldsymbol  z=[z_1^\top,\ldots, z_k^\top,\ldots,z_m^\top]^\top=\bar H\mathring {\boldsymbol p}$,  there exists a positive constant $\epsilon$ such that $\|z_k\|=\|\mathring p_i-\mathring p_j\|\ge \epsilon, \ \forall k \in \{1,\ldots, m\}$. That is, the bearing information $\mathring {\bar g}_k=\frac{z_k}{\|z_k\|} $ is well defined $\forall k \in \{1,\ldots, m\}$.
	
 	Now to prove the 'if' part of the item we use the fact that there exists at least one bearing vector $\bar g_q,\ q\in\{1,\ldots,m\}$ which is PE. This implies that there exist two constants $T>0$, $\mu_q >0$ such that $\forall t\ge0$ and for all fixed $\mathring {\boldsymbol p}\in S$ leading to $\boldsymbol z=\bar H \mathring {\boldsymbol p}$, we have
 	\begin{equation}
 	\begin{aligned}
 	\frac 1 T	\boldsymbol z^\top \int_t^{t+T}\Pi(\tau)d\tau \boldsymbol z=&\frac 1 T\sum_{k=1}^m  z_k^\top \int_t^{t+T}\pi_{\bar g_k(\tau)}d\tau z_k\\ \ge& \mu_q\|z_q\|^2.
 	\end{aligned}
 	\end{equation}
 	Choose $0<\mu <\mu_q\frac{\|z_q\|^2}{\|\boldsymbol z\|^2}$, we can get $$\frac 1 T \mathring {\boldsymbol p} ^\top \bar H^\top \int_t^{t+T}\Pi(\tau)d\tau \bar H\mathring {\boldsymbol p}\ge \mu \mathring {\boldsymbol p} ^\top \bar H^\top\bar H\mathring {\boldsymbol p}$$ which implies that $L_B(\boldsymbol p(t))$ is PE.
	
 	To prove the 'only if' part, we proceed hereafter by contradiction. Assume that none of the bearing vectors is PE which implies that for all $\mu_k>0$, $\forall T>0$, $\exists t\ge0$ and $\exists\boldsymbol z=\bar H \mathring {\boldsymbol p}$, such that $\frac 1 T z_k^\top \int_t^{t+T}\pi_{\bar g_k(\tau)}d\tau z_k< \mu_k\|z_k\|^2,\ \forall k\in\{1,\ldots,m\}$. Since $L_B(\boldsymbol p(t))$ is PE, there exists $T>0$ and $\mu >0$ such that, $\forall t\ge0$ and $\forall \boldsymbol z=\bar H \mathring {\boldsymbol p}$, $\frac 1 T \boldsymbol z^\top \int_t^{t+T}\Pi(\tau)d\tau \boldsymbol z\ge \mu\|\boldsymbol z\|^2$. Choose $\mu_k\le \frac{\mu\|\boldsymbol z\|^2}{m\|z_k\|^2}$, one concludes that, $\exists t>0$ and $\exists \boldsymbol z=\bar H \mathring {\boldsymbol p}$
 	\begin{equation}\scalebox{0.9}{$
 		\frac 1 T \boldsymbol z^\top \int_t^{t+T}\Pi(\tau)d\tau \boldsymbol z=\frac 1 T \sum_{k=1}^m z_k^\top \int_t^{t+T}\pi_{\bar g_k(\tau)}d\tau z_k < \mu\|\boldsymbol z\|^2$}
 	\end{equation}
 	which yields a contradiction.
	
 	  Proof of Item (3):
 %
 	Let $L_B^{'}(\boldsymbol p(t)^{'})=\bar H_1 ^{\top}\Pi_1\bar H_1$ be the bearing Laplacian matrix for the formation $\mathcal G^{'}(\boldsymbol p^{'}(t))$, where $\Pi_1=\diag(\pi_{\bar g_{k}}),k\in\{1,\ldots,m^{'}\}$ and $\bar H_1\in \mathbb R^{dm^{'}\times d(n-1)}$ is a submatrix of $\bar H=\begin{bmatrix}\bar H_1 & \boldsymbol 0_{m^{'}}\\h & u\end{bmatrix}$ where $u=\boldsymbol  1_{m-m^{'}}\otimes I_d$ and $h\in \mathbb R^{d\times d(n-1)}$ is the $\{0,-1\}$-matrix with rows indexed by edges and the columns by vertices: $[h]_{ki}=-1$ if vertex $i$ is associated to the edge $k$, and $[h]_{ki}=0$ otherwise $(k\in\{m^{'}+1,\ldots,m\}$ and $ i\in\{1,\ldots,n-1\})$. Hence $L_B(\boldsymbol p(t))=\bar H ^{\top}\Pi\bar H=\begin{bmatrix}
 	\bar H_1 ^{\top} \Pi_1\bar H_1 +h^\top \Pi_2 h& h^\top \Pi_2  u\\ u^\top \Pi_2 h &  u^\top \Pi_2  u
 	\end{bmatrix}$ with $\Pi_2=\diag(\pi_{\bar g_{k}}),k\in\{m^{'}+1,\ldots,m\}$. For any $\boldsymbol x=[\boldsymbol x'^\top,x_n^\top]\in \mathbb S^{dn-1}$, with $\boldsymbol x^{'}=[x_1^\top,\ldots,x_{n-1}^\top]^\top\in \mathbb R^{d(n-1)}$ and $\|x'\|\leq 1$, such that $ \bar H \boldsymbol x\ne 0$, one has
 	\begin{equation}\label{eq:xLB}
 	\boldsymbol x^\top L_B \boldsymbol x=\boldsymbol x^{'\top} L_B^{'}\boldsymbol x^{'}+(h\boldsymbol x^{'}+ux_n)^\top\Pi_2(h\boldsymbol x^{'}+ux_n).
 	\end{equation}	
 In order to prove the 'if' part, recall first that if $\mathcal G^{'}(\boldsymbol p^{'})$ is BPE, one has according to \eqref{eq:xLB} that  $\exists T>0, \mu_1>0$, $\forall t\ge0$, $\boldsymbol x^\top \frac 1 T\int_{t}^{t+T}L_B(\tau )d\tau \boldsymbol x\ge\mu_1\boldsymbol x^{'\top} \bar H_1^{\top}\bar H_1\boldsymbol x^{'}>0$ with $\bar H_1\boldsymbol x^{'}\ne 0$. When $\bar H_1\boldsymbol x^{'}= 0$, one has  $x_1=x_2=\ldots,=x_{n-1}=x_0$, with $x_0 \ne x_n$ to ensure that $ \bar H \boldsymbol x\ne 0$.  This in turn implies that $\boldsymbol x^\top \frac 1 T\int_{t}^{t+T}L_B(\tau )d\tau \boldsymbol x=(h\boldsymbol x^{'}+ux_n)^\top \frac 1 T\int_{t}^{t+T}\Pi_2(\tau)d\tau(h\boldsymbol x^{'}+ux_n)=(x_0-x_n)^\top\int_t^{t+T}\sum_{m^{'}+1}^m\pi_{\bar g_k(\tau)}d\tau (x_0-x_n)$. Now, using the fact that $\sum_{k=m^{'}+1}^m\pi_{\bar g_k}$ is PE one can ensure that there exists a $\mu_2>0$ such that $\boldsymbol x^\top \frac 1 T\int_{t}^{t+T}L_B(\tau )d\tau \boldsymbol x \ge \mu_2\|x_0-x_n\|^2>0$.
 From there, one concludes that $\exists T>0$ such that $\forall t\ge0$, $\boldsymbol x^\top \frac 1 T\int_{t}^{t+T}L_B(\tau )d\tau \boldsymbol x\ge\mu \boldsymbol x ^\top\bar H^\top \bar H \boldsymbol x$ with positive $\mu$ such that
 	\begin{equation*}\scalebox{0.9}{$
 	\mu= \left\{
 	\begin{aligned}
 		& \mu_1\frac{\|\bar H \boldsymbol x\|^2}{\|\bar H_1\boldsymbol x^{'}\|^2}, \text{ if } \bar H_1\boldsymbol x^{'}\ne0,\\
 		& \mu_2 \frac{\|\bar H \boldsymbol x\|^2}{\|x_0-x_n\|^2}, \text{ if } \bar H_1\boldsymbol x^{'}=0, \; (x_1=\ldots=x_{n-1}=x_0\ne x_n).
 	\end{aligned}\right.$}
 	\end{equation*}

	 	 For the 'only if' part, we proceed using a proof by contradiction. Since the formation $\mathcal G(\boldsymbol p(t))$ is BPE, $\forall t\ge0$, one has
 	 $\forall \boldsymbol x$ such that $\bar H \boldsymbol x\ne 0$, $\exists \mu>0, T>0$, $\boldsymbol x^\top \frac 1 T\int_{t}^{t+T}L_B(\tau )d\tau \boldsymbol x\ge \mu \boldsymbol x^\top \bar H ^\top \bar H \boldsymbol x$. If we assume that the matrix $\sum_{k=m^{'}+1}^m\pi_{\bar g_k}$ is not PE, then by  choosing $\boldsymbol x=[\boldsymbol 0,x_n]^\top$ (i.e. $\boldsymbol x^{'}=\boldsymbol 0_{d(n-1)} $ and $x_n\ne 0$), one has $\forall \mu >0$, $\forall T>0$, $\exists t\ge 0$, such that $\boldsymbol x^\top \frac 1 T\int_{t}^{t+T}L_B(\tau )d\tau \boldsymbol x=x_n^\top \int_t^{t+T}\sum_{k=m^{'}+1}^m \pi_{\bar g_k(\tau)}d\tau x_n <\mu x_n^\top x_n=\mu \boldsymbol x^\top \bar H ^\top \bar H \boldsymbol x$ which yields a contradiction.
 	\end{pf}
 Fig. \ref{fig:pe_edges3D}-\scalebox{0.8}{$(a1),(b1),(c1),(d1), (e1)$} and Fig. \ref{fig:pe_edges3D}-\scalebox{0.8}{$(a2),(b3),(c2),(d3), (e3)$} illustrate some examples for item (1) and (2) of Theorem \ref{thm:min_pe}, respectively. Combining Item (3) with Lemma \ref{lem:Q_norm}, one can easily explain the construction of Fig.\ref{fig:pe_edges3D}-$(d2)$ from $(c2)$ by adding one PE bearing and the construction of Fig.\ref{fig:pe_edges3D}-$(d3)$ from $(c2)$ by adding two non-collinear non-PE bearings. As for Item (3), it can be considered as a generalization of the vertex addition method defined in \cite{eren2007using} in which only static bearings are involved in a bearing-based Henneberg construction.

Now it is useful to keep in mind that the concept of BPE formation implies that a number of inter-agent bearings are time-varying and 
hence the formation is time-varying, possibly with time-varying shape and scale. A particular case arises when the whole formation is subjected to a similarity transformation involving a time-varying rotation. In this case, the shape is maintained constant (and possibly the scale also) but one can still guarantee that the formation is BPE, by considering that
$\forall t\ge0,\forall i\in \mathcal V$,
\begin{equation}\scalebox{0.95}{$
 p_i(t)=s(t)R(t)^\top p_i(0)+c(t), \label{sim}$}
 \end{equation}
 with $s(t)\in \mathbb R^+$, $c(t)\in \mathbb R^d$ and $R(t)\in SO(d)$ an orientation matrix of a virtual frame attached to some point on the formation with respect to the common inertial frame. This is what we term \textit{Relaxed Bearing Rigid} property for BPE formations, which differs from and is stronger than the classical bearing rigid concept since not only shape but also the scale of the formation  is constrained.
\begin{definition}
	A formation $\mathcal G(\boldsymbol p(t))$ is called \textit{Relaxed Bearing Rigid} (RBR), if it is BPE and subjected to the similarity transformation \eqref{sim}.
\end{definition}
\begin{proposition}
	Consider a formation $\mathcal G(\boldsymbol p(t))$ defined in $\mathbb{R}^{d}$ subjected to the similarity transformation \eqref{sim}. Then $\mathcal G(\boldsymbol p(t))$ is BPE if there exists a spanning three in $\mathcal G$ such that all the bearings  $\bar g_k(t)=R(t)^\top \bar g_k(0),\forall t\ge0$ associated to this spanning tree are PE.
\end{proposition}
 \begin{pf}
 From \eqref{sim}, it is straightforward to verify that $\bar g_k(t)=R(t)^\top \bar g_k(0),\forall t\ge0, \forall k\in\{1,\ldots,m\}$ with $R(t)\in SO(d)$ the orientation matrix of the similarity transformation.
 Consider now only a spanning three part of the graph involving the $n$ nodes and the $(n-1)$ PE bearings $g_k,\forall k\in\{1,\ldots,n-1\}$. Let $\bar H_s\in \mathbb R^{d(n-1)\times dn}$ denote the incidence matrix associated to the spanning tree which can be obtained by deleting the $n$th to $m$th rows of $\bar H$ and define $\Pi_s=diag(\pi_{\bar g_k})\in\mathbb R^{d(n-1)\times d(n-1)}, k\in\{1,\ldots,n-1\}$. From there, one ensures that $\bar H^\top \Pi \bar H\ge H_s^\top \Pi_s \bar H_s$, $\Null(\bar H_s^\top \bar H_s)=\Null(\bar H^\top \bar H)$ and hence $\exists c>0$ such that $H_s^\top  \bar H_s \ge c \bar H^\top \bar H$. Since $\Pi_s$ is PE, there exists $\mu>0$ and $T>0$, $\forall t\ge0$, $\bar H^\top \frac 1 T\int_t^{t+T}\Pi(\tau) \bar H\ge\bar H_s^\top \frac 1 T\int_t^{t+T}\Pi_s(\tau) \bar H_s\ge \mu \bar H_s^\top\bar H_s\ge c \mu\bar H^\top \bar H $ which in turns implies that $\mathcal G(\boldmath p(t))$ is BPE.
 %
 	\end{pf}
	\section{Bearing-only formation control for single-integrator dynamics in $\mathbb R^d$}\label{sec:control}
	In this section we propose distributed control laws for a multi-agent system with single-integrator dynamics using only bearing measurements, which guarantee UE stabilization of the actual formation to the desired one up to a translational vector provided the desired formation is BPE.
	Consider the formation $\mathcal{G}(\boldsymbol p)$ defined in Section \ref{subsection:formation control}, where each agent $i\in\mathcal{V}$ is modeled as a single integrator with the following dynamics:
	\begin{equation} \label{eq:single integrator}
	\dot{p}_i=v_i
	\end{equation}
	where $p_i\in\mathbb{R}^d$ is the position of the $i$th agent  and $v_i\in\mathbb{R}^d$ is its velocity input, as previously defined, both expressed in a common inertial frame. Similarly, let $p_i^*(t)$ and $v_i^*(t)\in\mathbb{R}^d$ denote the desired position and velocity of the $i$th agent, respectively, and define the desired relative position vectors $p_{ij}^*$ and bearings $g_{ij}^*$, according to \eqref{eq:pij} and \eqref{eq:gij}, respectively. Let $\boldsymbol{p}^*(t)=[p_1^{*\top}(t),...,p_n^{*\top}(t)]^\top\in \mathbb{R}^{dn}$ be the desired configuration. Let $\{\bar p_{k}^*(t)\}_{k\in\{1,...,m\}}$ and  $\{\bar g_{k}^*(t)\}_{k\in\{1,...,m\}}$ be the set of all desired edge vectors and desired bearing vectors, respectively, under an arbitrary orientation of the graph. Consider the following assumption:
	
	\begin{assumption} \label{ass:construction}
		The sensing topology of the group is described by a undirected graph $\mathcal{G}(\mathcal{V},\mathcal{E})$ which has a spanning tree and each agent $i\in \mathcal V$ can measure the relative bearing vectors $g_{ij}$ to its neighbors $ j\in \mathcal{N}_i$.
		The desired velocities $v_i^*(t)$ and desired positions $p_i^*(t)$ ($i\in\mathcal V$) are chosen such that, for all $t\ge 0$,  $v_i^*(t)$ are bounded, the resulting desired bearings $g_{ij}^*(t)$ are well-defined and the desired formation $\mathcal G(\boldsymbol p^*(t))$ is BPE.
	\end{assumption}	
	
	For each agent $i\in\mathcal{V}$, define the position error $\tilde{p}_{i}:=p_{i}-p_{i}^*$ along with the following kinematics:
	\begin{equation} \label{eq:states_f1}
	\dot{\tilde{p}}_{i}=v_i-v_i^*,
	\end{equation}
	and consider the following control law
	\begin{equation}
	v_i=-k_p\sum_{j\in \mathcal{N}_i}\pi_{g_{ij}}p_{ij}^*
	+v_i^*, \label{eq:ui1}
	\end{equation}
	where $k_p$ is a positive gain.
	Let $\tilde{\boldsymbol p}:=\boldsymbol p-\boldsymbol p^*$ be the configuration error. Using the control law \eqref{eq:ui1} for $i\in \mathcal V$, one gets:
	\begin{equation}\label{tildep}
	\dot{ \tilde{\boldsymbol p}}=-k_pL_B(\boldsymbol p(t))\tilde{\boldsymbol p}.
	\end{equation}
	
	\begin{thm} \label{thm:1st}
		Consider the error dynamics \eqref{eq:states_f1} along with the control law \eqref{eq:ui1}.
		If Assumption \ref{ass:construction} is satisfied, then, under any initial condition satisfying 
		\begin{equation}\label{eq:bound-p0}\scalebox{0.95}{$
		   \| \tilde {\boldsymbol p}(0)\|< \frac 1 {2}\min_{(i,j)\in\mathcal E}\|p_i^*(t)-p_j^*(t)\|,$}
		\end{equation}
		the feedback control law \eqref{eq:ui1} is well defined for all $t\ge 0$ and the following assertions hold
		\begin{enumerate}
			\item the relative centroid vector $q_0:=\frac{1}{n}U^\top \tilde {\boldsymbol p}(t)\in \mathbb R^{d}$ is invariant, that is, $q_0(t)= \frac{1}{n}U^\top \tilde {\boldsymbol p}(0)$;\\
			\item the equilibrium  $\tilde{\boldsymbol p}(t)-Uq_0=0$ is UE stable.
		\end{enumerate}
	\end{thm}
\vspace{-0.5cm}
	\begin{pf}
		We begin by assuming that the controller \eqref{eq:ui1} is well defined and then (in proof of Item 2) we show that it is well defined for all time.\\
		\textit{Proof of Item 1):}
		Since  $\Span\{U\}\subset\Null(L_B(\boldsymbol p(t)))$, it is straightforward to verify that:
		\begin{equation}
		\frac{d}{dt} q_0 =U^\top \dot {\tilde{\boldsymbol p}}/n=-\frac{k_p}{n}U^\top L_B(\boldsymbol p(t))\tilde{\boldsymbol p}\equiv 0, \label{eq:dotscale}
		\end{equation}
		and hence one concludes that  the relative centroid $q_0$ is constant ($q_0=\frac{1}{n}U^\top \tilde {\boldsymbol p}(0)=\frac1n \sum_{i\in\mathcal{V}} \tilde p_i(0))$.\\
		\textit{Proof of Item 2):}
		Define a new variable $\boldsymbol \delta:=\tilde{\boldsymbol p}-U q_0$
		and note that
		$\tilde{\boldsymbol p}$ can be decomposed into the following two orthogonal components
		\begin{align*}
		\tilde{\boldsymbol p} &= (I - \frac1n U U^T) \tilde{\boldsymbol p} + \frac1n U U^T\tilde{\boldsymbol p}
		=\boldsymbol \delta + U q_0.
		\end{align*}
		Since $U^TL_B=0$ and $L_B U=0$,  $\dot {\boldsymbol \delta}(t)=-L_B(\boldsymbol p(t))\boldsymbol \delta$.
		Considering the storage function
		$	\mathcal{L}_1=\frac 1 2 \|\boldsymbol \delta\|^2$,
	one can conclude that the time derivative of $\mathcal{L}_1$
		\begin{equation}
		\begin{aligned}
		\dot{\mathcal{ L}}_1 &=-k_p\boldsymbol \delta^\top L_B(\boldsymbol p(t))\boldsymbol \delta \leq 0
		\end{aligned}\label{eq:dotL1_semi}
		\end{equation}
		is negative semi-definite and $\boldsymbol \delta(t)$ is bounded and non-increasing for all $t\ge 0$, due to the fact that
		$L_B(\boldsymbol p(t))\geq 0$.
		Since $\boldsymbol \delta(t)$ and $Uq_0$ are orthogonal, it follows that
		$\| \tilde {\boldsymbol p}(t) \|^2 = \|\boldsymbol \delta (t)\|^2 +  \|U(\frac{1}{n}U^\top\tilde {\boldsymbol p}(0)) \|^2 \leq \| \tilde {\boldsymbol p}(0)\|^2$
		for all $t \ge 0$.
		
		In order to show that $g_{ij}(t),\forall (i,j)\in \mathcal E$ is well defined $\forall t\ge 0$ which in turn implies that \eqref{eq:ui1} is well defined under the proposed initial condition, we have to show that $p_{ij},\forall (i,j)\in \mathcal E$ never crosses zero. Using the fact that $p_{ij}=\tilde p_j-\tilde p_i+p_{ij}^*$, one gets
		\begin{equation}\label{eq:bound_pij}\scalebox{0.85}{$
			\begin{aligned}
			 \|p_{ij}(t)\|\ge \|p_{ij}^*(t)\|-\|\tilde p_i(t)\|-\|\tilde p_j(t)\|\ge\|p_{ij}^*(t)\|-2\|\tilde {\boldsymbol p}(t)\|.
			\end{aligned}$}
		\end{equation}
	Combining this with \eqref{eq:bound-p0} and the fact that $\|\tilde {\boldsymbol p}(t)\|\le\|\tilde {\boldsymbol p}(0)\|,\forall t>0$, one concludes that $p_{ij}(t)\ne 0,\forall t$. 
	
		As for the proof of the UE stable of the equilibrium point $\boldsymbol \delta =0$ we recall
		that \eqref{eq:dotL1_semi} can be rewritten as
		\begin{equation}\scalebox{0.95}{$
		\begin{aligned}
		\dot{\mathcal{ L}}_1 &=-k_p\tilde{\boldsymbol p}^\top L_B(\boldsymbol p(t))\tilde{\boldsymbol p}=-k_p\sum_{k=1}^m \bar p_k^{*\top}\pi_{\bar g_{k}}\bar p_k^*\\
		&=-k_p\sum_{k=1}^m\frac{\|\bar p_k^*\|^2}{\|\bar p_k\|^2}(\bar p_k-\bar p_k^*)^{\top}\pi_{\bar g_{k}^*}(\bar p_k-\bar p_k^*)
		\end{aligned}\label{eq:dotL1re}$}
		\end{equation}
		Using the fact that $\|\bar p_k(t)\|\le\|\bar p_k^*(t)\| + 2\| \tilde {\boldsymbol p}(0)\|$ $<\max\|\bar p_k^*(t)\| + \min\|\bar p_k^*(t)\|$ 
		(obtained analogously to \eqref{eq:bound_pij} and in combination with \eqref{eq:bound-p0}),  
		one gets
		\begin{equation}
		\begin{aligned}
		\dot{\mathcal{ L}}_1 &\le-\gamma\boldsymbol \delta^\top L_B(\boldsymbol p^*(t))\boldsymbol \delta,
		\end{aligned}
		\end{equation}
		with 
		$\gamma= k_p\left( \frac{\min\|\bar p_k^*(t)\|}{\max\|\bar p_k^*(t)\| +\min\|\bar p_k^*(t)\|}\right)^2>0.$
		Since $U^\top \boldsymbol \delta(t)\equiv 0$ and the desired formation is BPE, one can ensure that
		\begin{equation*}\scalebox{0.95}{$
		\begin{aligned}
		\frac 1 T \boldsymbol \delta(t)^\top \int_t^{t+T}L_B(\boldsymbol p^*(\tau))d\tau\boldsymbol \delta(t)&\ge\mu \boldsymbol \delta(t)^\top\bar H^\top \bar H \boldsymbol \delta(t)\\ &\ge\mu \lambda_{dn-d} \|\boldsymbol \delta(t)\|^2,
		\end{aligned}$}
		\end{equation*}
		recall that $\lambda_{dn-d}$ is the smallest positive eigenvalue of $\bar H ^\top \bar H$. Hence, condition (1) of Theorem \ref{thm:ES} is satisfied. Since $L_B(\boldsymbol p(t))$ is bounded and condition (2) of Theorem \ref{thm:ES} is also satisfied, one concludes that $\boldsymbol \delta=0$ is UE stable.		
		\begin{remark}
			Note that although the closed-loop dynamics \eqref{tildep} is
			similar to the observer error dynamics in the proof of Theorem \ref{thm:1st} in \cite{tang2021relaxed}, 
			only local exponential stability can be ensured here while global
			exponential convergence of the observer error is guaranteed. For both
			cases the bearing Laplacian is the same but the PE conditions are not. For the observer design it is assumed
			that the actual formation is BPE while for controller design the BPE is
			assumed for the desired formation.
			The latter condition does not guarantee that the actual bearings, the
			bearing Laplacian, and hence the control law are well-defined for all
			time, since collisions may occur during the time evolution of the
			formation. This in turn implies that the actual state of the formation
			will always admit an exception set of critical points that cannot be
			part of the basin of attraction of the desired equilibrium. Theorem
			\ref{thm:1st} provides a conservative estimate for the basin of
			attraction, corroborating the idea that if the initial conditions are
			sufficiently close to a desired formation that is well-defined for all
			time then no collisions will occur and exponential convergence is
			guaranteed.
			
		\end{remark}
	\begin{remark}
		If $\dot {g}^{*}_{ij}=0$, $ v_{ij}^*=0, \forall (i,j)\in \mathcal E$ and the desired bearing formation is IBR, the proposed control laws \eqref{tildep} are the same as those proposed by \cite{zhao2016bearing} and hence the regulation of the actual formation to the desired one can be ensured up to a translation Euclidean vector and an additional scale factor.
	\end{remark}
	\end{pf}
	\section{Bearing formation control for double-integrator dynamics in $\mathbb R^d$}\label{sec:control2}
	In this section we extend the bearing formation control law for a multi-agent system with double-integrator dynamics in $\mathbb R^d$.
	Consider the formation $\mathcal{G}(\boldsymbol p)$ defined in Section \ref{subsection:formation control}, where each agent $i\in\mathcal{V}$ is modeled as a double integrator with the following dynamics:
	\begin{equation} \label{eq:double integrator}
	\left\{
	\begin{aligned}
	\dot{p}_i&=v_i\\
	\dot v_i&=u_i,
	\end{aligned}
	\right.
	\end{equation}
	where $u_i\in\mathbb{R}^d$ expressed in a common inertial frame is the acceleration input. Let $u_i^*(t)\in\mathbb{R}^d$ denote the desired acceleration of the $i$th agent and $\boldsymbol v^*(t)=[v_1^{*\top}(t),...,v_n^{*\top}(t)]^\top\in \mathbb{R}^{dn}$ the stacked velocity vector of the desired configuration $\boldsymbol p^*(t)$.
	Hereafter, the following assumption is made.
	\begin{assumption} \label{ass:construction2}
		The sensing topology of the group is described by an undirected graph $\mathcal{G}(\mathcal{V},\mathcal{E})$ which has a spanning tree and each agent $i\in \mathcal V$ can measure its velocity $v_i$ and the relative bearing vectors $g_{ij}$ to its neighbors $ j\in \mathcal{N}_i$.
		The desired acceleration $u_i^*(t) $, desired velocity $v_{i}^*(t)$, and desired position $p_{i}^*(t)\ (i\in \mathcal V)$ are chosen such that $u_i^*(t)$ and $v_{i}^*(t)$ are bounded, the resulting desired bearings $g_{ij}^*(t)$ are well-defined and the desired formation $G(\boldsymbol p^*(t))$ is BPE, for all $t\ge 0$.
	\end{assumption}
%
	
	For each agents $i\in\mathcal{V}$, define the velocity error $\tilde{v}_{i}:=v_{i}-v_{i}^*$. Then the error dynamics of error states $(\tilde{p}_{i},\tilde{v}_{i})$ can be represented as:
	\begin{equation} \label{eq:states_f}
	\left\{
	\begin{aligned}
	\dot{\tilde{p}}_{i}&=\tilde{v}_i\\
	\dot {\tilde{v}}_i&=u_i-u_i^*.
	\end{aligned}
	\right.
	\end{equation}
	The following control law is proposed for each agent $i\in\mathcal{V}$
	\begin{equation}
	u_i=-k_p\sum_{j\in \mathcal{N}_i}\pi_{g_{ij}}p_{ij}^*-k_d\tilde{v}_i
	+u_i^* \label{eq:ui}
	\end{equation}
	where $k_p$ and $k_d$ are positive constant gains. Defining the new variable $\tilde{\boldsymbol v}:=\boldsymbol v-\boldsymbol v^*$ and under the control law \eqref{eq:ui}, the dynamics of $(\tilde {\boldsymbol p},\tilde {\boldsymbol v})$ can be presented as
	\begin{equation} \label{eq:dynamics}
	\left\{
	\begin{aligned}
	\dot{\tilde{\boldsymbol p}}&=\tilde{\boldsymbol v}\\
	\dot {\tilde{\boldsymbol v}}&=-k_pL_B(\boldsymbol p(t)) \tilde{\boldsymbol p}-k_d\tilde{\boldsymbol v}
	\end{aligned}
	\right.
	\end{equation}
	\begin{thm} \label{lem:2agent}
		Consider the error dynamics \eqref{eq:states_f} along with the control law \eqref{eq:ui}.
		If Assumption \ref{ass:construction2} is satisfied and the positive gains $k_d$ and $k_p$ are chosen such that $k_d>\frac{k_p}{4}n(n-1)+1$, then for any initial condition such that
		\begin{equation}\label{bass}
		\begin{aligned}
		\| [\tilde{\boldsymbol p}(0)^\top \tilde{\boldsymbol v}(0)^\top]\|
		< \frac 1 {2b}\min_{(i,j)\in\mathcal E}\|p_i^*(t)-p_j^*(t)\|,
		\end{aligned}
		\end{equation}
		with $b=\max\{\sqrt{\frac{\lambda_{\max}(P)}{\lambda_{\min}(P)}},\sqrt 2\}$ and $P=\frac 1 2 \begin{bmatrix}
		k_d I_{dn}    &  I_{dn} \\
		I_{dn}& I_{dn}\\
		\end{bmatrix}>0$,
		the feedback control \eqref{eq:ui} is well defined and the following two assertions hold $\forall t\ge 0$:
		\begin{enumerate}
			\item  the relative centroid and its velocity $(q_0(t), \dot{q}_0(t) )=(\frac{U^\top\tilde {\boldsymbol p}(t)}n,\frac{U^\top\tilde {\boldsymbol v}(t)}n) \in \mathbb R^{2d}$ converge UE to $(\mathring{q_0}=q_0(0)+\frac 1 {k_d}\dot{q}_0(0), 0$),\\
			\item  the equilibrium $(\tilde{\boldsymbol p}-U\mathring{q}_0,\tilde{\boldsymbol v})=(0,0)$  is UE stable.
		\end{enumerate}
	\end{thm}
	\begin{pf}
		Analogously to the proof of Theorem \ref{thm:1st}, we assume first that the controller \eqref{eq:ui} is well defined and then we show that is it so in the proof of Item 2.\\
		\textit{Proof of Item 1):}
		From \eqref{eq:dynamics} and due to the fact that $\Span\{U\}\subset\Null(L_B(\boldsymbol p(t)))$, one has:
		$\ddot q_0(t)=-k_d \dot q_0(t)$
		which implies that $\dot q_0(t)=\dot q_0(0)\exp(-k_d t)$ and $q_0(t)=\frac{1}{k_d}( 1-\exp(-k_dt))\dot q_0(0)+q_0(0)$. From \eqref{bass}, one concludes that $(q_0(t), \dot q_0(t))$ converges UE to $(\mathring q_0,0)$.\\
		\textit{Proof of Item 2):} Similarly to the proof of Theorem \ref{thm:ES}, we define $\tilde {\boldsymbol x}:=[ (\tilde {\boldsymbol p}-Uq_0)^\top, ( \tilde {\boldsymbol v}-U\dot{q}_0)^\top]^\top$ and note that $[(U q_0)^\top\ (U\dot q_0)^\top] \tilde{\boldsymbol x} = 0$, meaning that
		$[\tilde {\boldsymbol p}(t)^\top, \tilde {\boldsymbol v}(t)^\top] = \tilde {\boldsymbol x}(t)^\top + [(U q_0(t))^\top\ (U\dot q_0(t))^\top]$ and the two components are orthogonal. We will first show that $\tilde {\boldsymbol x}$ 
		is bounded. Using \eqref{eq:dynamics}, it is straightforward to verify that:
		\begin{equation}
		\dot {\tilde {\boldsymbol x}}(t)=-A(t)\tilde {\boldsymbol x}(t) \label{eq:dotx}
		\end{equation}
		with $A=$\scalebox{0.9}{$\begin{bmatrix}
		0_{dn}    & -I_{dn} \\
		k_pL_B&k_dI_{dn}\\
		\end{bmatrix}$}.
		Considering the following positive definite storage function
		$\mathcal{L}_2=\tilde {\boldsymbol x}^\top P \tilde {\boldsymbol x}$,
		one can verify that
		\begin{equation}
		\begin{aligned}
		\dot{\mathcal{ L}}_2 &=-\tilde {\boldsymbol x}^\top Q(t)\tilde {\boldsymbol x},
		\end{aligned}\label{eq:dotL_semi}
		\end{equation}
		with $Q(t)=$\scalebox{0.9}{$\begin{bmatrix}
		k_pL_B(\boldsymbol p(t))     & \frac {k_p} 2 L_B(\boldsymbol p(t)) \\
		\frac {k_p} 2 L_B(\boldsymbol p(t))&(k_d-1) I_{dn}\\
		\end{bmatrix}$}.
		The matrix $Q$ can be decomposed as $Q=S^\top M_Q S$ with $S=$\scalebox{0.9}{$\begin{bmatrix}
		\Pi \bar H  & 0_{dm\times dn}\\
		0_{dn}&I_{dn} \\
		\end{bmatrix}$} and $M_Q=$\scalebox{0.9}{$\begin{bmatrix}
		k_pI_{dm}& \frac {k_p} 2\bar H\\ \frac {k_p} 2\bar H^\top & (k_d-1)I_{dn}
		\end{bmatrix}$}, which shows that if $k_d>\frac{k_p}4 \|\bar{H}\|^2+1$ then $Q\ge 0$ and $\dot{\mathcal{ L}}_2$ is negative semi-definite.
		Since $ n(n-1) \geq \tr(\bar{H}^T\bar{H})\geq\|\bar{H}\|^2$, 
		one concludes that $\tilde {\boldsymbol x}(t)$ is bounded which in turn implies that $\tilde{\boldsymbol p}$ is bounded. Since $\tilde {\boldsymbol x}$ and $[(Uq_0)^\top \ (U\dot q_0)^\top]^\top$ are orthogonal, $q_0(t)=\frac{1}{k_d}( 1-\exp(-k_dt))\dot q_0(0)+q_0(0)$ and $k_d>1$, $\tilde {\boldsymbol p}(t)$ can be bounded by
		\begin{equation}\scalebox{0.8}{$\label{eq:norm_p}
			\begin{aligned}
			&\|\tilde {\boldsymbol p}(t)\|^2=\|\tilde {\boldsymbol p}(t)-Uq_0(t)\|^2+\|Uq_0(t)\|^2\\
			&\le \|\tilde {\boldsymbol x}(t)\| +  2(\frac{1}{k_d^2}\|Uq_0(0)\|^2 + \|U\dot q_0(0)\|^2)\\
			&\le \frac{\lambda_{\max}(P)}{\lambda_{\min}(P)}\|\tilde {\boldsymbol x}(0)\|^2 + 2(\|Uq_0(0)\|^2 + \|U\dot q_0(0)\|^2)\le b^2\| [\tilde{\boldsymbol p}(0)^\top \tilde{\boldsymbol v}(0)^\top]\|^2
			\end{aligned}$}
		\end{equation}
		recalling that $b=\max\{\sqrt{\frac{\lambda_{\max}(P)}{\lambda_{\min}(P)}},\sqrt 2\}$.

		To prove that \eqref{eq:ui} is well defined, it suffices to show that $p_{ij},\forall (i,j)\in \mathcal E$ never crosses zero. We proceed analogously  to  the proof of item 2 - Theorem \ref{thm:1st}. Using \eqref{eq:bound_pij}, \eqref{bass} together with \eqref{eq:norm_p}, one concludes that:
		$ \forall t \ge 0,\| p_{ij}(t)\|\ge\| p_{ij}^*(t)\| - 2b \| [\tilde{\boldsymbol p}(0)^\top \tilde{\boldsymbol v}(0)^\top]\|> 0$.

		Now, to prove that $(\tilde {\boldsymbol p}-U\mathring q_0, \tilde {\boldsymbol v}(t))=(0,0)$ is also UE stable,  it suffices to prove that $\tilde {\boldsymbol x}=0$ is UE stable.
		Since $Q=S^\top M_Q S$, $k_d>\frac{k_p}{4}\|\bar H\|^2+1$, and $\bar p_k^{*\top}\pi_{\bar g_k}\bar p_k^*=\frac{\|\bar p_k^*\|^2}{\|\bar p_k\|^2}\bar p_k^\top\pi_{\bar g_k^*} \bar p_k$, one has
		\begin{equation*}\scalebox{0.95}{$
		\dot{ \mathcal L}_2=-\tilde {\boldsymbol x}^\top S^\top M_Q S \tilde {\boldsymbol x}\le -\lambda_o\tilde {\boldsymbol x}^\top S^\top S \tilde {\boldsymbol x}\le-\gamma\tilde {\boldsymbol x}^\top \Sigma\tilde {\boldsymbol x}\le 0$}
		\end{equation*}
		where \scalebox{0.9}{$\lambda_o\geq\left(\frac{2dn-1}{\tr(M_Q)}\right)^{2dn-1} \det(M_Q)>0$} (\cite{MERIKOSKI1997101})
		and \scalebox{0.9}{$\Sigma(t)=\begin{bmatrix}  L_B(\boldsymbol p^*(t)) & 0_{dn} \\ 0_{dn} &I_{dn}\end{bmatrix}$}.
		Since $\mathcal{L}_2$ is non-increasing, one ensures that $\|\bar p_k(t)\|\leq \| \bar p_k^*(t)\| + 2b \| [\tilde{\boldsymbol p}(0)^\top \tilde{\boldsymbol v}(0)^\top]\|$ (analogously to the single integrator case) and together with  \eqref{bass}, one obtains
		\begin{equation*}\scalebox{0.8}{$
		\begin{aligned}
		 \gamma=\lambda_o\min(\frac{\|\bar p_k^*(t)\|}{\|\bar p_k(t)\|})^2=\lambda_o\left(\frac{\min\|\bar p_k^*(t)\|}{\max\|\bar p_k^*(t)\| +\min\|\bar p_k^*(t)\|}\right)^2>0
		\end{aligned}$},
		\end{equation*}
		 which is independent of the initial conditions. 
		Using the BPE condition of the desired formation and the fact that $\Span\{U\}=\Null(L_B)$, one concludes that condition (1) in Theorem \ref{thm:ES} is satisfied. By direct application of Lemma \ref{lem:c}, condition (2) in Theorem \ref{thm:ES} is also satisfied, and therefore $\tilde {\boldsymbol x}=0$ is UE stable. This in turn implies that $(\tilde{\boldsymbol p}-U\mathring{q}_0,\tilde{\boldsymbol v})=(0,0)$  is UE stable.	
	\end{pf}
	\section{Simulation Results}\label{sec:sim}
In this section, simulation results are provided to validate the controllers for multi-agent system under both single- and double- integrator dynamics.

\begin{figure}[!htb]
	\centering
	\includegraphics[scale = 0.7]{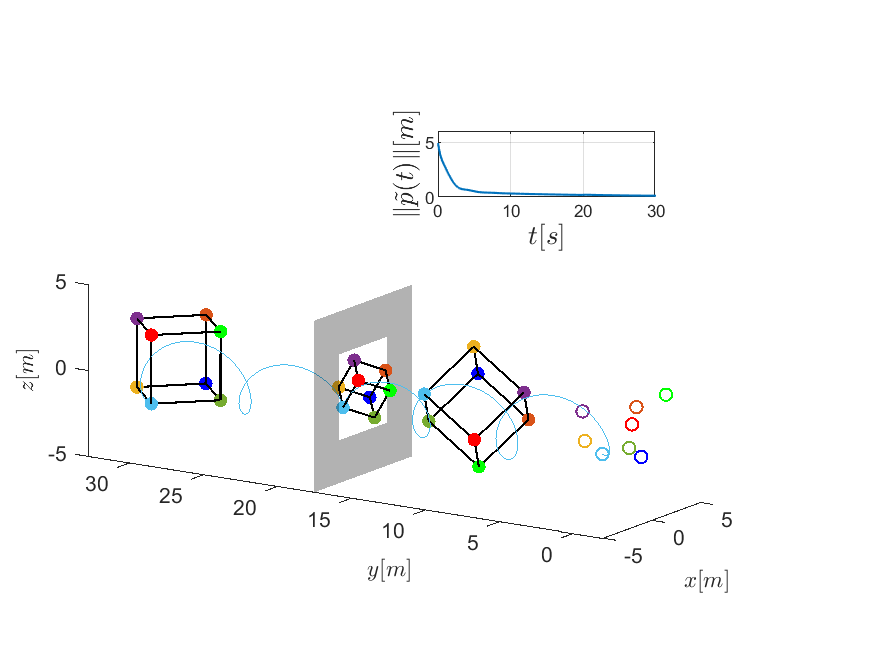}	
	\caption{The figure shows four snapshots of the 3-D evolution of a cubic formation: $t=0$, the initial conditions (empty circles); $t=10$s, when agents are converging to the desired formation; $t=17.5$s, when agents are passing through a gap with reduced scale; $t=30$s, when agents already passed the gap. The blue line represents the trajectory of the blue agent and the black lines represent the connections between agents.}
	\label{fig:3D}
\end{figure}

For the single integrator dynamics system, we consider a 8-agent system in 3-D space. The desired formation is chosen such that $p_i^* (t) = r(t)R(t)^\top p_i^*(0)+[0 \ t/5 \ 0]^\top$, with $r(t)$ a time-varying scale, \scalebox{0.9}{$R(t)=\begin{bmatrix}
	1 &0& 0\\ 0&\cos(\frac {\pi} {3}t)& -\sin(\frac {\pi} {3}t) \\0& \sin(\frac  {\pi} {3}t) &\cos(\frac  {\pi} {3}t)
	\end{bmatrix}$}, $p_1^*(0)=[\sqrt{2}\ 0 \ -1]^\top,p_2^*(0)=[0\ \sqrt{2}\ -1 ]^\top, p_3^*(0)=[-\sqrt{2}\ 0 \ -1]^\top, p_4^*(0)=[0 \ -\sqrt{2}\  -1]^\top,p_5^*(0)=[\sqrt{2}\ 0 \ 1]^\top,p_6^*(0)=[0\ \sqrt{2}\ 1 ]^\top, p_7^*(0)=[-\sqrt{2}\ 0 \ 1]^\top$, and $p_8^*(0)=[0 \ -\sqrt{2}\  1]^\top$, which form a cube in $\mathbb R^3$ that rotates about the $x$-axis and translates along $y$-axis as show in Fig. \ref{fig:3D}. Note that the desired formation is not IBR but RBR. The initial conditions are chosen such that $q_0=0$ (the initial centroid coincides with the initial centroid of the desired formation). 
The chosen gain is $k_p=1$. The bottom of Fig. \ref{fig:3D} shows the evolution of the formation in three dimensional space and the top shows the evolution of the error variable $\tilde {\boldsymbol p}(t)$. One can see that the formation converges to the desired one after $t=10s$ and the desired scale is time-varying such that the desired cubic formation passes through a narrow gap at $t=17.5$. We can conclude that, under the proposed bearing-only control laws, the formation achieves the desired geometric pattern in terms of shape and scale without the need for bearing rigidity nor any distance between two agents. What's more, if one of the agents is assigned to be the leader, the formation tracking problem can be solved without  imposing initial conditions of $q_0=0$, hence the task of collision avoidance such as passing through a narrow passage can be accomplished.

\begin{figure}[!htb]
	\centering
	\includegraphics[scale = 0.8]{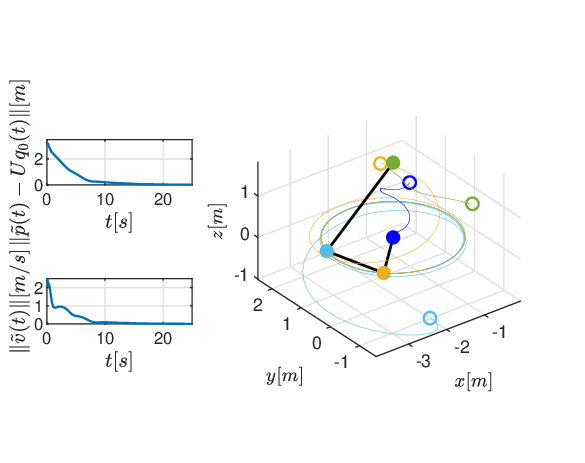}	
	\caption{Time evolution of the relative position error ($\|\tilde{\boldsymbol p}(t)-U q_0(t)\|$) and velocity error ($\|\tilde{\boldsymbol v}(t)\|$) (left hand side) and 3-D trajectories (right hand side) for a RBR pyramid formation: the empty and solid circles represent the initial and final positions of the agents, respectively. The colored lines are the trajectories of the agents and the black lines are the connections between the agents. }
	\label{fig:3D_2nd}
\end{figure}

For the multi-agent system under double-integrator dynamics, we consider a RBR desired formation with the graph topology that has only one spanning tree, in which the four agents form a pyramid shape in $\mathbb{R}^3$ that rotates about one of the agents (Fig. \ref{fig:3D_2nd}). The desired positions of the agents are such that $p_i^* (t) = R(t)^\top p_i^*(0)$, with \scalebox{0.9}{$R(t)=\begin{bmatrix}
	\cos(\frac {\pi} {4}t)& -\sin(\frac {\pi} {4}t) &0\\ \sin(\frac  {\pi} {4}t) &\cos(\frac  {\pi} {4}t)& 0\\0 &0& 1
	\end{bmatrix}$}, $p_1^*(0)=[0\ 0\ 0]^\top,p_2^*(0)=[1\ 0\ 0 ]^\top, p_3^*(0)=[0.5\ -\sqrt{3}/2\ 0]^\top$ and $p_4^*(0)=[\sqrt{3}/2 \ -0.5 \ 1]^\top$.  The right hand side of Fig.~\ref{fig:3D_2nd} shows the time evolution of the 3-D formation converging to the desired one and left hand side shows the time evolution of error states $\tilde{\boldsymbol p}(t)-U q_0(t)$ and $\tilde {\boldsymbol v}(t)$, respectively. It validates the fact that the proposed control laws stabilize the formation without requiring bearing rigidity. Additional animations can be found in \url{https://youtu.be/lAtphz1mBfQ}.
	\section{Conclusion}\label{sec:conc}
	This paper presents new results on formation control of both kinematic and dynamic systems based on time-varying bearing measurements. The key contribution is to show that if the desired formation is BPE,
	relaxed conditions on the interaction topology (which do not require bearing rigidity) can be used to derive distributed control laws that guarantee uniform exponential stabilization of the desired formation only up to a translation vector. Simulations results are provided to illustrate the performance of the proposed control method. Future work will focus on redesigning the distributed control laws to actively provide inter-agent collision avoidance.
	\begin{ack}    
	\footnotesize{                          
		This work was partially supported by the Project MYRG2018-00198-FST of the University
		of Macau; by the Macao Science and Technology, Development Fund under Grant FDCT/0031/2020/AFJ; by Funda\c{c}\~{a}o  para a Ci\^{e}ncia e a Tecnologia (FCT) through Project UID/EEA/50009/2019 and Project PTDC/EEI-AUT/5048/2014;
		and by the ANR-DACAR project. The work of Z. Tang was supported by FCT through Ph.D. Fellowship PD/BD/114431/2016 under the FCT-IST NetSys Doctoral Program.}  
	\end{ack}
	\bibliographystyle{agsm}
	\scriptsize{\bibliography{bibliography}}
	\appendix
	\section{Technical Lemmas and Theorem }  
	\small{
%
	\begin{thm}\label{thm:ES}
		Consider the following system
		\begin{equation}\label{dot x}
		\dot x(t)=f(x(t),t),\ x \in \mathbb{R}^{dn}
		\end{equation}
		with $f(x(t),t)$ a piecewise continuous and locally Lipschitz function such that $f(0,t)=0$. Assume there exists a function $\mathcal L_x(t)=\mathcal L (t,x(t))\in \mathbb R^+$, such that $\lambda_1\|x(t)\|^2\le \mathcal{L}_x(t)\le \lambda_2\|x(t)\|^2$ and $\dot{\mathcal{L}}_x(t)\le -\gamma x(t)^\top \Sigma(t) x(t)$, where $\Sigma(t)\in \mathbb{R}^{dn\times dn}$ is an upper bounded  positive semi-definite function $(\|\Sigma(t)\|\le \lambda_{\Sigma})$, with $\lambda_1$, $\lambda_2$, $\lambda_{\Sigma}$ and $\gamma$ positive constants. If $\forall t\ge 0$,
		\begin{enumerate}
			\item $\exists T>0,\ \exists\mu>0$, $ \frac 1 Tx(t)^\top\int_t^{t+T}\Sigma(\tau)d\tau x(t)\ge \mu \|x(t)\|^2$, and
			\item $\dot {\mathcal{L}}_x(t)\le -\frac 1 c \|f(x,t)\|^2\le 0$, $c>0$,
		\end{enumerate}
		then the origin of \eqref{dot x} is UE stable, and verifies: $x(t)\le \sqrt{\frac{\lambda_2}{\lambda_1(1-\sigma)}}x(0)\exp(-\frac \sigma {2T} t)$ with $\sigma = \frac{1}{1+\rho}\frac{1}{1+ \rho c T^2\gamma\lambda_\Sigma}$ and $\rho = \frac{\lambda_2}{\mu T\gamma}$.
	\end{thm}
	\begin{pf}
		 The proof follows the arguments used in \cite[Lemma 5]{loria2002uniform}. Taking integral of  $\dot{\mathcal{L}}_x(t)\le -\gamma x(t)^\top \Sigma(t) x(t)$, we get
		\begin{equation}\scalebox{1}{$ \label{eq:intL21}
			\mathcal{L}_x(t+T)-\mathcal{L}_x(t)\le-\gamma\int_{t}^{t+T}\|\Sigma^{\frac1 2}(\tau){x}(\tau)\|^2 d \tau$}
		\end{equation}
		where, according to \eqref{dot x}, $x(\tau)$ can be rewritten as
		\begin{equation}\scalebox{1}{$\label{int_x21}
			x(\tau)=x(t)+\int_{t}^{\tau}f(x(s),s)ds.$}
		\end{equation}
		To obtain a bound for the integral term in \eqref{eq:intL21}, we substitute \eqref{int_x21} in $\|\Sigma^{\frac1 2}(\tau){x}(\tau)\|^2$ and use $\|a+b\|^2\ge[\rho/(1+\rho)]\|a\|^2-\rho \|b\|^2$ and Schwartz inequality to obtain
		\begin{equation} \label{sigma_bound}\scalebox{0.9}{$
			\|\Sigma^{\frac1 2}(\tau){x}(\tau)\|^2\ge\frac{\rho}{1+\rho} \|\Sigma^{\frac1 2}(\tau){x}(t)\|^2 -\rho\lambda_{\Sigma}T\int_t^\tau \|f(x(s),s)\|^2ds.$}
		\end{equation}
		Substituting \eqref{sigma_bound} into  \eqref{eq:intL21}, we obtain
		\begin{equation}\scalebox{0.9}{$
			\begin{aligned}
			&\mathcal{L}_x(t+T)-\mathcal{L}_x(t)\le
			-\frac{\gamma\rho}{1+\rho}\int_{t}^{t+T}\|\Sigma^{\frac 1 2}(\tau)x(t)\|^2d\tau\\
			&+\rho \gamma\lambda_{\Sigma} T
			\int_t^{t+T}\int_t^{\tau}\|f(x(s),s)\|^2ds d\tau.
			\end{aligned}$}
		\end{equation}
		Using the condition (1) and (2), we have
		\begin{equation}\scalebox{0.85}{$
			\begin{aligned}\label{eq:intL}
			&\mathcal{L}_x(t+T)-\mathcal{L}_x(t)\le
			-\frac{\mu T \gamma\rho}{(1+\rho)}\|x(t)\|^2-c\rho \gamma\lambda_{\Sigma}T
			\int_t^{t+T}\int_t^{\tau}\dot{\mathcal{L}}(s)ds d\tau.
			\end{aligned}$}
		\end{equation}
		Changing the order of integration in equation \eqref{eq:intL}, one can get
		\begin{equation}\scalebox{0.85}{$
			\begin{aligned} \label{eq:change of int}
			-
			\int_t^{t+T}\int_t^{\tau}\dot{\mathcal{L}}_x(s)ds d\tau
			\le-T\int_t^{t+T}\dot{\mathcal{L}}_x(s)ds= T(\mathcal{L}_x(t)-\mathcal{L}_x(t+T))
			\end{aligned}$}	
		\end{equation}
		Substituting inequality \eqref{eq:change of int} into \eqref{eq:intL} we have
		\begin{equation*}\scalebox{0.85}{$
			\begin{aligned}
			\mathcal{L}_x(t+T)\le(1-\sigma)\mathcal{L}_x(t),\ \sigma:=\frac{\rho {\mu T \gamma}}{(1+\rho)(1+\rho cT^2\gamma\lambda_{\Sigma} )\lambda_2}.
			\end{aligned}$}	
		\end{equation*}
		By choosing $\rho = \frac{\lambda_2}{\mu T \gamma}$, one has $\sigma = \frac{1}{1+\rho}\frac{1}{1+ \rho c T^2\gamma\lambda_\Sigma} < 1$. For any $t\ge 0$, let $N$ be the smallest positive integer such that $t\le NT$. Since $\mathcal{L}_x(t)\leq \mathcal{L}_x((N-1)T)\leq(1-\sigma) \mathcal{L}_x((N-2)T)$, $\mathcal{L}_x(t)$ can be bounded by a staircase geometric series such that
		$\mathcal{L}_x(t) \leq (1-\sigma)^{N-1} L_x(0)$ and hence the exponential convergence follows from
		$\mathcal{L}_x(t) \leq (1-\sigma)^{N-1} \mathcal{L}_x(0)=\frac{\exp(-bNT)}{1-\sigma} \mathcal{L}_x(0)\le\frac{\exp(-bt)}{1-\sigma} \mathcal{L}_x(0) $
		with
		$b = \frac{1}{T}\ln(\frac{1}{1-\sigma}) > \frac \sigma T.$		
	\end{pf}
	\begin{lemma} \label{lem:c}
		Consider the matrix $A$ and $Q$ defined in equation \eqref{eq:dotx} and equation \eqref{eq:dotL_semi} respectively. Assume $k_d>\frac{k_p}{4}\|\bar H\|^2+1$. There exists $c>0$ such that $cQ-A^\top A\geq 0$.
	\end{lemma}
\vspace{-0.5cm}
	\begin{pf}
		Define \scalebox{0.9}{$S=\begin{bmatrix}
			\Pi \bar H  & 0_{dm\times dn}\\
			0_{dn}&I_{dn} \\
			\end{bmatrix}$}, \scalebox{0.9}{$M_Q=\begin{bmatrix}
			k_pI_{dm}& \frac {k_p} 2\bar H\\ \frac {k_p} 2\bar H^\top & (k_d-1)I_{dn}
			\end{bmatrix}$} and \scalebox{0.9}{$M_A=\begin{bmatrix}
			k_p^2\bar H\bar H^\top &k_p k_d\bar H\\ k_p k_d \bar H^\top & (1+k_d^2)I_{dn}
			\end{bmatrix}$}, then $Q=S^\top M_Q S$ and $A^\top A=S^\top M_A S$.
		We can conclude that $cQ-A^\top A\ge 0$, if $cM_Q-M_A\ge0$
		which holds if $c$ is chosen such that $(k_pk_d-k_p-\frac{k_p^2}{4}\|\bar H\|^2)c^2-(k_pk_d^2+k_p-k_p^2\|\bar H\|^2)c+k_p^2\|\bar H\|^2\ge 0$ and $c\ge \max\{k_p\|\bar H\|^2,\frac{k_d^2+1}{k_d-1}\}$.
		
	\end{pf}
}

\end{document}